\documentclass[reprint,prd,superscriptaddress,nofootinbib]{revtex4-1}
\usepackage{color}
\usepackage{epsfig}
\usepackage{amsmath,bm}
\usepackage{blindtext}
\usepackage[modulo]{lineno}
\usepackage{xspace}
\newcommand{\nldbd}{\ensuremath{0\nu\beta\beta}\xspace}
\newcommand{\mlightest}{\ensuremath{m_1}\xspace}
\newcommand{\mbb}{\ensuremath{m_{\beta\beta}}\xspace}

\begin{document}
\title{Discovering neutrinoless double-beta decay in the era of precision neutrino cosmology}

\author{Manuel Ettengruber}\email{manuel@mpp.mpg.de}
\affiliation{Max-Planck-Institut f\"ur Physik (Werner-Heisenberg-Institut), F\"ohringer Ring 6, 80805 M\"unchen, Germany}

\author{Matteo Agostini}\email{matteo.agostini@ucl.ac.uk}
\affiliation{Department of Physics and Astronomy, University College London, Gower Street, London WC1E 6BT, UK}

\author{Allen Caldwell}\email{caldwell@mpp.mpg.de}
\affiliation{Max-Planck-Institut f\"ur Physik (Werner-Heisenberg-Institut), F\"ohringer Ring 6, 80805 M\"unchen, Germany}

\author{Philipp Eller}\email{philipp.eller@tum.de}
\affiliation{Technical University Munich (TUM), James-Franck-Strasse 1, 85748 Garching, Germany}

\author{Oliver Schulz}\email{oschulz@mpp.mpg.de}
\affiliation{Max-Planck-Institut f\"ur Physik (Werner-Heisenberg-Institut), F\"ohringer Ring 6, 80805 M\"unchen, Germany}

\begin{abstract}
     We evaluate the discovery probability of a combined analysis of proposed neutrinoless double-beta decay experiments in a scenario with normal ordered neutrino masses. The discovery probability strongly depends on the value of the lightest neutrino mass, ranging from zero in case of vanishing masses and up to 80-90\% for values just below the current constraints. We study the discovery probability in different scenarios, focusing on the exciting prospect in which cosmological surveys will measure the sum of neutrino masses. Uncertainties in nuclear matrix element calculations partially compensate each other when data from different isotopes are available. Although a discovery is not granted, the theoretical motivations for these searches and the presence of scenarios with high discovery probability strongly motivates the proposed international, multi-isotope experimental program. 
\end{abstract}

\date{\today}
\maketitle
Neutrinoless double-beta (\nldbd) decay is a lepton-creating nuclear transition
in which two neutrons simultaneously convert into two protons and two electrons~\cite{Agostini:2022zub}. 
This nuclear decay would change the difference between the number of leptons and antileptons ($L$), while preserving the difference between the number of baryons and antibaryons ($B$). Processes changing $B-L$ are not foreseen in our standard model of particle physics and have never been observed, but their existence is required by our best theories explaining why the universe contains much more matter than antimatter~\cite{Fukugita:1986hr}. The discovery of \nldbd\ decay would not only provide the first direct observation of a process violating $B-L$, it would also prove Majorana's hypothesis that neutrinos are their own antiparticles~\cite{Majorana:1937vz,Racah:1937qq,Schechter:1981bd}. Majorana's neutrinos would get their mass differently from any other fermion, and their apparently unnaturally small values could be understood within models where the neutrino masses are inversely proportional to those of heavy right-handed partners~\cite{Minkowski:1977sc,GellMann:1980vs,yanagida:1979as,Mohapatra:1979ia}.
At present, the search for \nldbd\ decay is our most sensitive test for $B-L$ violating physics and Majorana's neutrino masses, which could both be connected to the same new physics at ultrahigh-energy scales.

Growing interest in \nldbd-decay experimental searches also comes from their interplay with cosmology. Such an interplay was already present between the experiments conducted in the last decade---see for instance Ref.~\cite{DellOro:2016tmg}---but it will become far more important in the next few years, as discussed in this work.
Indeed, neutrinos deeply affect both Big Bang nucleosynthesis and the large scale structure of the universe. In particular, they induce characteristic signatures in the relative abundance of elements as well as in the power spectra of the cosmic microwave background (CMB) and baryon acoustic oscillations (BAO)~\cite{Lattanzi:2017ubx}.
These effects can be used to set upper bounds on the sum of neutrino masses ($\Sigma=m_1+m_2+m_3$), and the current best results indicate $\Sigma < 120$\,meV (95\% credible interval), driven by the measurements from Planck and its combination with lensing and BAO data~\cite{Aghanim:2018eyx}. 
The next generation surveys, DESI~\cite{Font-Ribera:2013rwa} and EUCLID~\cite{Euclid:2021qvm}, promise to measure $\Sigma$ with 20\,meV precision even assuming its minimally allowed value. A future measurement of $\Sigma$ would set a clear target for the \nldbd-decay half-life, creating an exciting synergy between these two fields.  With DESI already taking data and EUCLID starting operation next year, a measurement of $\Sigma$ could be announced at any time. 

The \nldbd-decay half-life strongly depends on the particle physics process expected to mediate the decay, which could be 
Majorana neutrinos or other new BSM physics. In this work, we focus on the exchange of light Majorana neutrinos interacting via standard, weak left-handed currents.
This mechanism is very popular as it can take place already in a minimal extension of the standard model in which neutrinos are massive Majorana fermions. In addition, it is typically the dominant mechanism even in more complex models in which multiple channels are allowed. In this scenario, the half-life of the decay is given by:
\begin{equation}
 \frac{1}{T_{1/2}} = G\, g_A^4  \, \mathcal{M}^2 \left( \frac{m_{\beta \beta}}{m_e} \right)^2,
 \label{halflife}
\end{equation}
where $G$ is the kinematically allowed phase space factor, $g_A^4\simeq1.276$ is the axial-vector coupling,  $\mathcal{M}$ the nuclear matrix element (NME) accounting for the overlap between the nucleon wave functions in mother and daughter isotopes, and $m_e$ the electron mass.
The effective Majorana mass \mbb\ expresses the contribution of the three virtual neutrinos mediating the decays and the probability for a neutrino interacting as a right-handed chiral state. It is defined as:
\begin{equation}
    \mbb = |c_{12}^2 c_{13}^2m_1 + s_{12}^2c_{13}^2m_2 e^{i \alpha_1} + s_{13}^2m_3 e^{i \alpha_2}|,
    \label{mbetabeta}
\end{equation}
where $c_{ij}$ and $s_{ij}$ are the cosines and sines of the lepton-mixing angles, $m_i$ the eigenvalues of the neutrino mass eigenstates and $\alpha_{i}$ are the so-called Majorana phases~\cite{Zyla:2020zbs}. 

Our capability to predict the decay half-life is limited by two main factors. The first one is related to the precision and accuracy of the many-body calculations used to estimate the NME values. 
Four primary many-body methods have been historically used in the field: the nuclear shell model (NSM), the quasiparticle random-phase approximation (QRPA) method, energy-density functional (EDF) theory, and the interacting boson model (IBM). Several calculations per method are available, each characterized by different assumptions and approximations. Their results can differ by up to a factor of three for a given isotope, and significant differences are present even within each method~\cite{Agostini:2022zub}. 

The second factor limiting the accuracy of our predictions is the value of \mbb.
Indeed, although neutrino oscillation parameters have been accurately measured, we currently have no information on the Majorana phases and the value of the lightest neutrino mass eigenstate~\cite{Zyla:2020zbs}. We also do not know the ordering of the neutrino mass eigenstates. 
Global fits~\cite{Esteban:2020cvm} currently show a mild preference for the normal ordering, but its significance is still under debate~\cite{Jimenez:2022dkn, Gariazzo:2022ahe}. Furthermore,
cosmological bounds on the sum of the neutrino masses disfavor parts of the available parameter space for inverted ordering, while the parameter space of normal ordering remains largely untouched. 
If neutrino masses follow the inverted ordering, \mbb\ is constrained and its minimally allowed value is $18.4 \pm 1.3$\,meV~\cite{Agostini:2021kba}.  Should neutrino masses follow a normal ordering, vanishing \mbb values are in principal possible, even if they require a precise tuning of the value of the Majorana phases resulting in the cancellation of the terms in Eq.~\ref{mbetabeta}~\cite{Feruglio02,Benato:2015via}.  Achieving a sensitivity to at least probe \mbb down to the minimum value allowed for the inverted ordering has been for two decades the holy grail of \nldbd-decay experiments. 

The search for \nldbd decay is at a turning point;
the community has developed experimental concepts to probe the full parameter space available for inverted ordering. A discussion is taking place in the community to define the next steps. 
As part of this process, the United States’ Department of Energy has recently carried out a ton-scale-experiment portfolio review, which led to a summit
involving the Astroparticle Physics European Consortium (APPEC), American and European funding agencies and the scientific community. 
Three experiments are already at the conceptual design stage and can be pushed forward:  CUPID~\cite{CUPID:2019imh}, LEGEND~\cite{LEGEND:2021bnm} and nEXO~\cite{nEXO:2018ylp}.
These experiments use different isotopes, and have the potential to perform independent and complementary measurements.
Having data from multiple isotopes is not only needed to corroborate a future discovery, but it will also boost the overall discovery power and reduce the impact of systematic uncertainties related to both the detection concept and the nuclear many-body calculations. It could also put light on the mechanism mediating the decay~\cite{Lisi:2022nka,Graf:2022lhj}.

In this work, we study the discovery prospect of the future, multi-isotope, global endeavour to discover \nldbd\ decay.  As a discovery is granted in case of inverted-ordered Majorana neutrinos, we focus on the discovery odds for normal-ordered neutrinos. We use all existing neutrino data to constrain \mbb\ and calculate Bayesian discovery probabilities for future searches under different scenarios. 
The crucial parameters in this kind of analysis are the Majorana phases and the value of the lightest mass eigenstate $m_1$. Their prior distributions strongly influence the results of the analysis. As in the approach suggested in ~\cite{Benato:2015via}, we express the lack of information on the phases by assuming a uniform prior distribution.  We do not see a reasonable alternative choice.  

We treat the prior choice $m_1$ differently from our previous work~\cite{Agostini:2017jim,Caldwell:2017mqu}, in that we first provide discovery odds as a function of $m_1$.  This makes manifest the strong dependence on this parameter. We then assume a flat prior on $m_1$ and consider scenarios in which cosmological constraints on $\Sigma$ give indirect information on it, reducing the influence of the prior choice. In particular, after considering the current constraints on $\Sigma$, we focus on the two most extreme hypothetical scenario in which DESI and EUCLID will measure 
$\Sigma=100\pm20$\,meV, which is just below the current limits, or
$\Sigma=59\pm20$\,meV, which is at the bottom of the expected parameter space allowing for $m_1\approx 0$\,meV with significant probability.  

\begin{figure}[tb]
    \centering
    \includegraphics[width=\columnwidth]{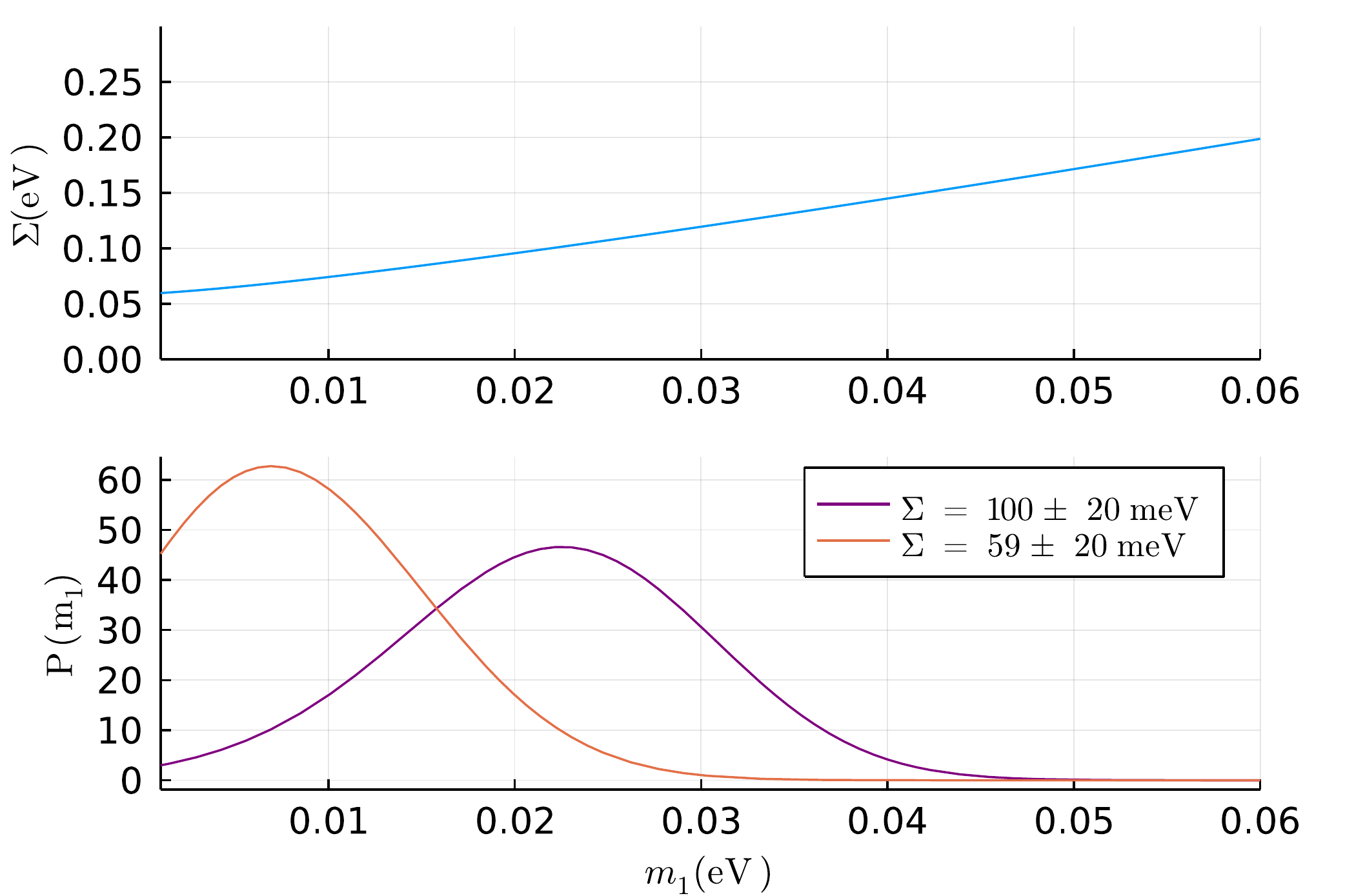}
    \caption{(Top) Correlation between $\Sigma$ and $m_1$ assuming the best fit values for the neutrino oscillation angles and mass splittings~\cite{Zyla:2020zbs}. Assuming neutrino masses follow the normal ordering, $\Sigma$ is constrained to be larger than 59\,meV. (Bottom) Gaussian probability distributions of $\Sigma$ transformed into probability distributions of $m_1$ through a change of variable using the best fit of the neutrino oscillation parameters. The Gaussian distributions $\Sigma=59\pm20$\,meV and $100\pm20$\,meV correspond to the two extreme measurements that DESI and EUCLID can perform.}
    \label{Sigma}
\end{figure}

When the oscillation parameters are fixed, $\Sigma$ and $m_1$ are connected by a bijective function and probability distributions can be analytically computed using a change of variable~\cite{DellOro:2019pqi,Agostini:2020oiv}.
For illustration, \figurename~\ref{Sigma} shows the probability distributions of $m_1$ corresponding to the two Gaussian probability distributions on $\Sigma$. The Jacobian of the transformation skew the distributions, creating tails on their left side and shifting their mode to larger values. 

In our analysis, we combined the likelihoods from the most sensitive \nldbd-decay experiments which are CUORE~\cite{CUORE:2021gpk}, EXO-200~\cite{EXO-200:2019rkq}, GERDA~\cite{GERDA:2020xhi},  and KamLAND-Zen~\cite{KamLAND-Zen:2022tow}. None of these have reported hints for a signal and have set lower limits on its half-life at the level of $10^{25} - 10^{26}$ years, corresponding to upper limits on \mbb of the order of 100\,meV. We also include the likelihood from the latest analysis of KATRIN~\cite{KATRIN:2021uub} on the electron neutrino mass 
$m_{\beta}= (c_{12}^2\,c_{13}^2\,m_1^2 + s_{12}^2\,c_{13}^2\,m_2^2 + s_{13}^2\,m_3^2)^{1/2}$. 
The parameters of interest for a \nldbd-decay analysis are collected in the vector $\theta$
\begin{equation}
    \theta = (\mlightest, \Delta m_{12}, \Delta m_{13}, s_{12}, s_{13}, \alpha_1, \alpha_2, \mathrm{NME})
    \label{theta}
\end{equation}
The oscillation parameters are incorporated into the analysis using Gaussian terms with central values and uncertainties taken from Ref.~\cite{Zyla:2020zbs}.

By sampling the likelihood function and prior probability distributions, we generate pseudo-data sets for the future \nldbd-decay experiments and evaluate their average probability to report a discovery, as proposed in Ref.~\cite{Caldwell:2006yj}. We reproduce the performance of future experiments by using a Poisson counting analysis with fixed background expectation as proposed in Ref.~\cite{Agostini:2022zub}, from which we also take the input effective background levels and signal efficiencies. We assume ten years of operation for all experiments, corresponding to what the community aims to achieve within the next two decades.
The discovery criteria is defined by requiring the posterior odds to be above a certain threshold, i.e.:
\begin{equation}
     \mathcal{O}_1 = \frac{P(D|H_1)}{P(D|H_0)} \frac{P(H_1)}{P(H_0)}>10,
    \label{BayesFactor}
\end{equation}
where $P(D|H)$ are the probabilities of the data given the hypothesis that \nldbd\ decay exists ($H_1$) or not ($H_0$).  $P(H_1)$ and $P(H_0)$ are their corresponding priors  assumed to be equal. This criteria corresponds to the request that $H_1$ is ten times more probable than $H_0$ assuming they are initially equally probable. We finally define as discovery probability the fraction of pseudo-datasets satisfying our discovery criteria.
Our calculations are performed using the BAT software kit and its native Metropolis-Hastings sampling algorithm~\cite{Schulz:2020ebm}.
We determined that the discovery criteria used in this work provides results numerically similar to those of a $3\sigma$ frequentist rejection test of $H_0$.
More details on our discovery probability calculations are given in the appendix.

We perform our calculations using fixed sets of NME values. We take each set from a specific many-body calculation, and consider calculations~\cite{Hyvarinen15,Simkovic18,Rodriguez10,Vaquero14,Song17,Barea15,Deppisch20}
whose results are available for all isotopes of interest in this analysis, i.e., $^{76}$Ge, $^{100}$Mo,$^{130}$Te, and $^{136}$Xe. This choice excludes some NSM and QRPA calculations for which the NME value for $^{100}$Mo is currently not available but it has the advantage that each element in a NME set has correlated systematic uncertainties that  partially cancel out when combining data on different isotopes \cite{Faessler:2011rv, Lisi:2022nka}.
The spread among discovery probabilities computed for different sets of NME values will hence give a rough idea of the uncertainty due to the different many-body methods. However, it will not capture effects coherently affecting all methods, such as the lack of the contact operator~\cite{Cirigliano19long} or the so-called ``$g_A$ quenching'' physics~\cite{Agostini:2022zub} that we discuss later.

\begin{figure}[tb]
  \centering
  \includegraphics[width=\columnwidth]{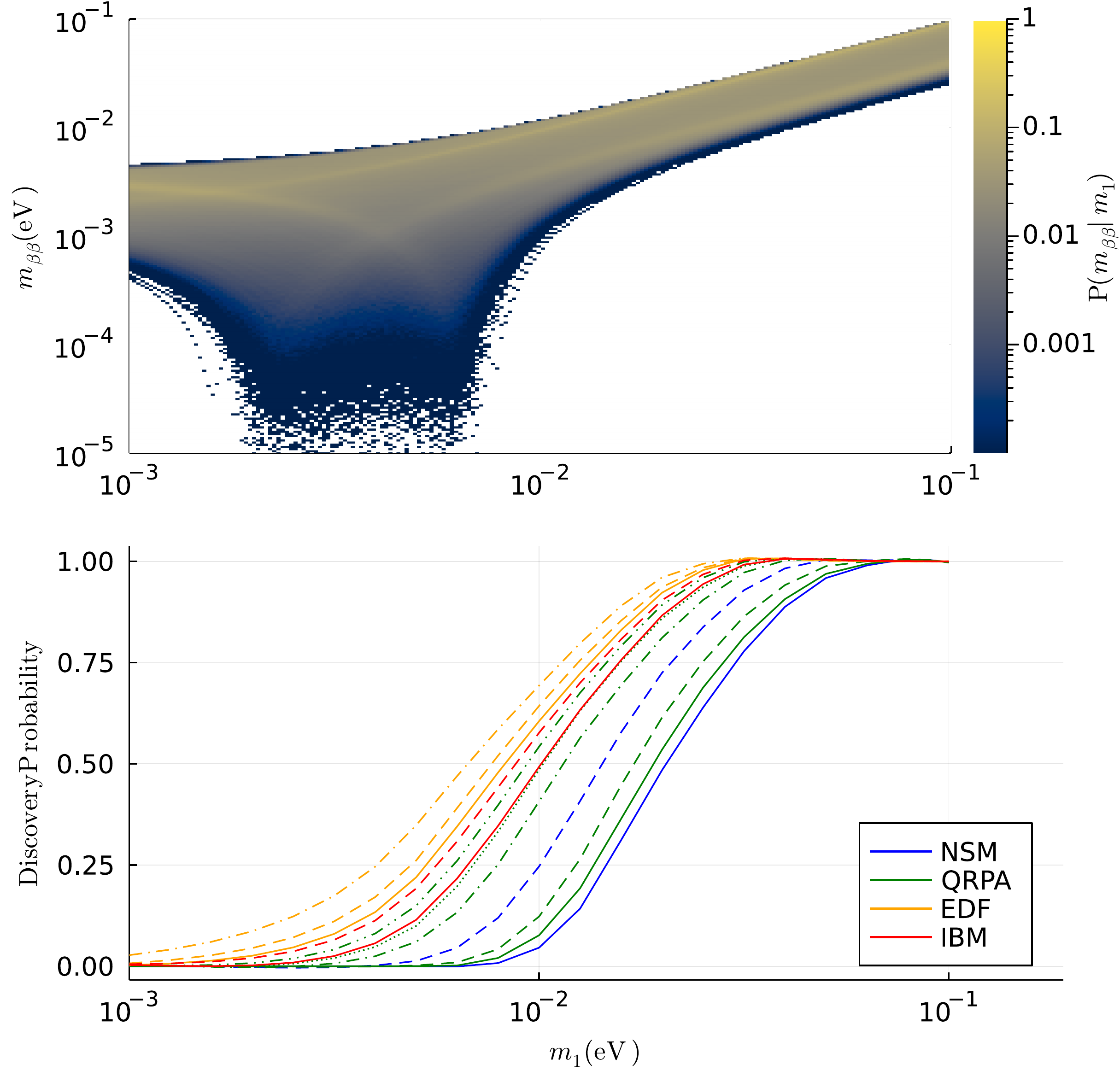}
  \caption{(Top) Conditional one-dimensional posterior probabilities for \mbb computed for a scan of fixed \mlightest values, assuming all available data on neutrinos, normal ordering and a uniform prior on the effective Majorana phases. (Bottom) Discovery probability for a combined analysis of CUPID, LEGEND and nEXO as a function of the true value of \mlightest. The probability is computed for different sets of NME values yielding a band for the discovery probability.}
  \label{Scan}
\end{figure}

The top panel of \figurename~\ref{Scan} shows the \mbb posterior probability distributions computed for a scan of fixed \mlightest values, ranging from $10^{-4}$ to $10^{-2}$\,eV. It should not be interpreted as a two-dimensional distribution, but rather as contiguous one-dimensional conditional probability distributions of \mbb, each normalized independently.
The probability distribution is contained in a well defined part of the parameter space thanks to the accurate measurements available for the neutrino oscillation parameters. The remaining width of \mbb probability distributions is due to the freedom left to the Majorana phases. Our choice of using a uniform prior for these parameters favors the largest \mbb values available at each fixed \mlightest value, including in the region between $10^{-3}-10^{-2}$\,eV where specific values of the Majorana phases can lead to vanishing \mbb values. The smaller is the value chosen for \mlightest, the smaller is the maximally-allowed value of \mbb, whose minimum reaches 0.5\,meV for $\mlightest=0$\,meV.
\nldbd-decay experiments cut into the upper part of the probability distributions, and are currently ruling out \mbb values above 156\,meV~\cite{KamLAND-Zen:2022tow}, indirectly constraining  \mlightest to be $\lesssim 100$\,meV.
Future experiments will reach discovery sensitivities down to \mbb\ values of 6\,meV depending on the NME values~\cite{Agostini:2021kba}.

The lower panel of \figurename~\ref{Scan} shows the combined discovery probability of CUPID, LEGEND and nEXO as a function of \mlightest for all sets of NME values considered.
The discovery probability starts at zero when \mlightest is smaller than 1\,meV, and continuously grows until it approaches 100\% when \mlightest is larger than $\sim$60\,meV. 
The discovery probability varies significantly depending on the considered set of NME values in the $m_1$ range 5--50\,meV. The larger the NME values, the higher are the discovery probabilities. The discovery probabilities converge below 5\,meV and above 50\,meV.

Given a theoretical prediction on the value of \mlightest, the discovery probabilities in the field of \nldbd-decay can be directly obtained from the plot in \figurename~\ref{Scan}. However, we are currently lacking a complete model of fermion masses and theory is not providing strong guidance on the value of \mlightest. For this reason, it is relevant to consider scenarios in which \mlightest is a free parameter weakly constrained by indirect information. The drawback of this approach is that, if the information on \mlightest is not strong enough, the results of any analysis are deeply affected by the choice of its prior probability distribution. In particular, any scale-invariant log-flat prior would lead to a non-normalizable posterior distribution unless a cut-off on \mlightest is applied as done in Ref.~\cite{Caldwell:2017mqu}. Other approaches effectively forcing the value of \mlightest to be similar to that of the other two mass eigenvalues have also been explored, see for instance~\cite{Agostini:2017jim, Simpson:2017qvj}. In the following, we consider a uniform prior distribution on \mlightest from 0 to 600\,meV. This prior choice favors \mbb values closer to the parameter space probed by the experiments if one has no other guidance of the parameter range of \mlightest. If, however, one includes cosmological bounds on $\Sigma$ the probability distribution for \mlightest is modified and analyses including cosmological data are less affected by the chosen prior on \mlightest.

\begin{figure}[tb]
  \centering
  \includegraphics[width=\columnwidth]{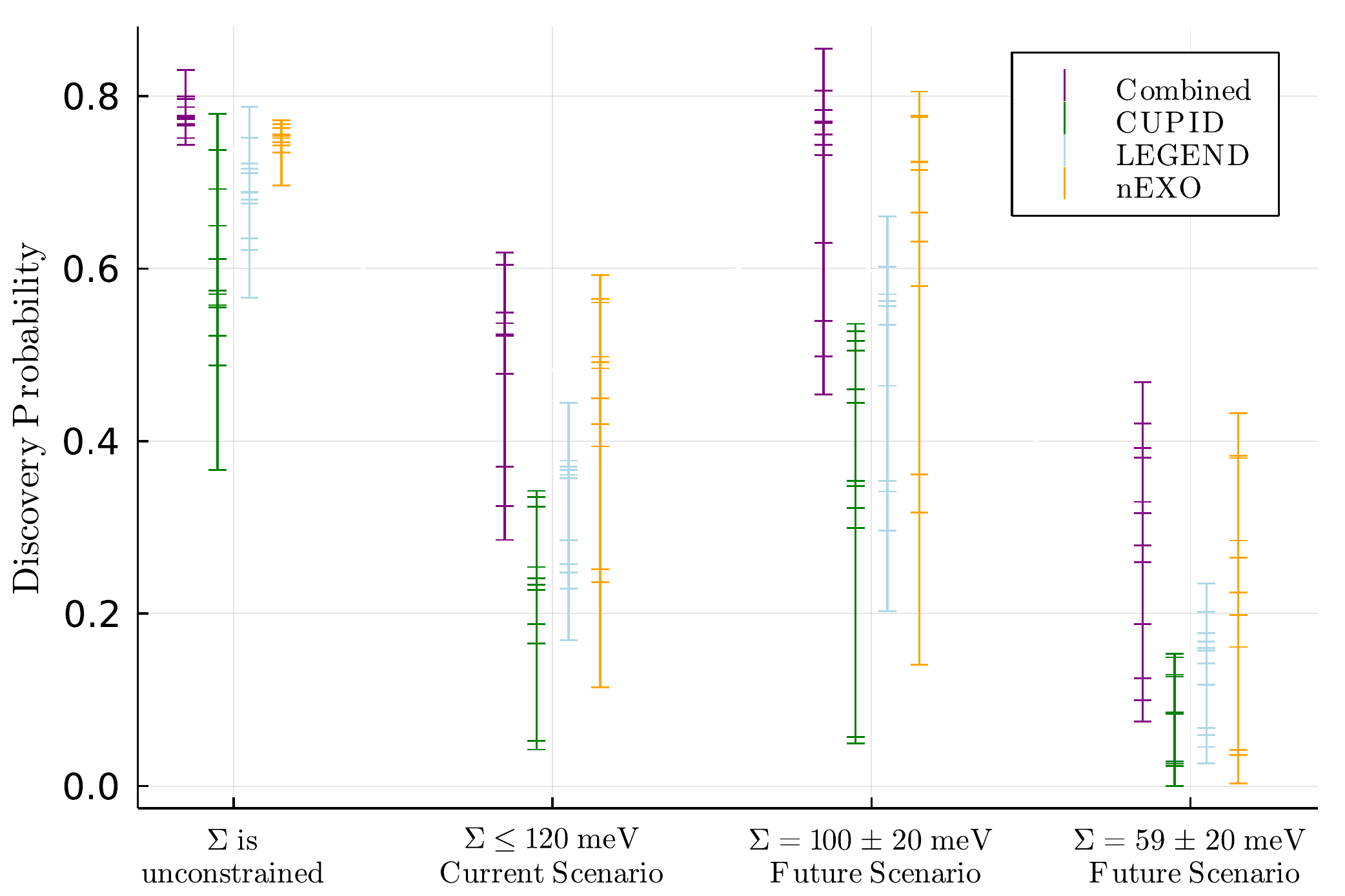}
  \caption{Discovery probabilities for a selection of proposed experiments and their combination under different scenarios and set of NME values. The scenarios differ because of the information included on $\Sigma$, respectively no information, current upper limit, and two possible measurements at the extreme of the currently allowed parameter space. The calculation has been performed using fixed sets of NME values, and each results is shown as an horizontal tick.}
  \label{Scenarios}
\end{figure}

\figurename~\ref{Scenarios} shows the discovery probabilities for CUPID, LEGEND, nEXO and their combination, under four scenarios and for each set of NME values. The first two scenarios show the impact of the use of the current cosmological constraint on $\Sigma$.
When considering cosmological models beyond $\Lambda_\textrm{CDM}$, much larger neutrino masses are allowed \cite{Alvey:2021xmq}, and the most stringent information on \mlightest comes from current \nldbd-decay experiments and from KATRIN. In such a scenario the discovery probabilities are as high as 80\% as the uniform prior on \mlightest has significant probability mass at larger $m_1$ values. In other words, if one ignores standard cosmological bounds and assumes a flat prior on $m_1$, the conclusion is that the discovery of \nldbd is quite probable. When the likelihood constraining $\Sigma<120$\,meV is included, this penalizes large \mlightest values reducing the discovery probability to values ranging between 20 and 60\%.

The speculative scenarios including future measurements of $\Sigma$ show encouraging discovery opportunities.
Should $\Sigma$ be right below the current constraints, for instance 100\,meV, future \nldbd-decay experiments are very likely to observe a signal, with discovery probability between 20 and 80\%. Even if $\Sigma=59 \pm 20$\,meV, which is at the bottom of its allowed parameter space, the discovery probabilities are significant, ranging from a few percents to above 40\%.
The discovery probabilities in these two scenarios are weakly affected by the prior choice on $m_1$, for which the measurements on $\Sigma$ gives robust information (see \figurename~\ref{Sigma}).
For instance, we estimated that a log-prior on \mlightest with a cutoff as in \cite{Caldwell:2017mqu} would reduce these discovery probabilities by a maximum of $30\%$.
This indicates that whatever value of $\Sigma$ will be reported by DESI and EUCLID, next-generation \nldbd-decay experiments will explore a complementary parameter space, where both the measurement of a signal or its exclusion will provide invaluable information.

The discovery probabilities of the single experiments are similar to each other, and the spread is maximal for nEXO and minimal for LEGEND, consistently with the spread of NME values available for Xe and Mo. 
When cosmological data are not used, \mlightest is primarily constrained by the current \nldbd-decay experiments. In this case, increasing the NME values pushes the \mbb probability distribution to lower masses, but also allows future experiments to probe lower \mbb values. The impact of such an interplay becomes negligible in the scenarios in which $\Sigma$ is constrained and the current \nldbd-decay experiments have a weak impact on the \mbb probability distribution.

The spread of results due to the NME uncertainty remains significant, especially considering that we did not include QRPA calculations for which $^{100}$Mo results are not available. A significant effort is ongoing within the nuclear theory community and will improve the accuracy and precision of NME values. 
{\it Ab initio} calculations have been performed for light and medium-sized nuclei~\cite{Gysbers:2019uyb}, and will soon be available for the heavier isotopes of interest. These new calculations are expected to incorporate more realistic nuclear correlations and corrections to the leading-order operator in chiral effective field theory, e.g. two-body currents, and first results suggest reduced NME values~\cite{Belley:2020ejd}. The inclusion of this so-called ``$g_A$ quenching'' physics can however be at least partially compensated by the previously neglected contact term recently introduced in Ref.~\cite{Cirigliano:2018hja}, leading to discovery probabilities similar to those shown in \figurename~\ref{Scenarios}. Indeed, we have computed that a 20\% overall scaling of the NME values for all isotopes (equivalent to a $\sim$10\% variation in $g_a$) would change the discovery probabilities by 5--10\%, not affecting the overall conclusions of our work. 
The impact of such an overall scaling is marginal as it consistently affects both the current and future experiments.

The combination of all experiments results in an average boost of discovery probabilities of about 20\% compared to the mean from single experiments. Additionally,
the range of values for the combination is significantly narrowed compared to the set of single experiments, as expected by the partial compensation of NME value fluctuations among the three isotopes.The combination mitigates the least favourable NME values leading
to the very small discovery probabilities in single experiments. An
other advantage of combining multiple experiments is an increased
confidence in a discovery. Indeed, systematic uncertainties related for
instance to miss-modelled background components will affect only a
single experiment and be mitigated by a combined analysis. Statistical
fluctuations will also compensate, providing a lower chance of false
discoveries which we estimated to be below the $0.2\%$ level. All these
arguments emphasize the value of executing several large-scale \nldbd
experiments.

In conclusion, precision neutrino cosmology and searches for \nldbd decay are heavily entangled and largely complementary. If the neutrino is a Majorana particle in the minimal extension of the Standard Model of particle physics, and the mass ordering is inverted, future \nldbd-decay experiments will clearly see a signal.  The situation for the normal ordering is more complicated, and results from future cosmological experiments will considerably narrow the allowed ranges for \mlightest and therefore the  discovery probability of next-generation \nldbd-decay experiments. If cosmology reports upper bounds on $\Sigma$ also in the future, there is moderate discovery probability for future \nldbd-decay searches and the question whether the neutrino is a Majorana particle will still be open. However, if cosmological results report a value for the sum of neutrino masses $\Sigma > 59$\,meV, the chance of discovering \nldbd-decay will be very significant also for the normal ordering.  A non-observation of \nldbd-decay in this case would give a strong indication that the neutrino is a Dirac particle.

\section{Acknowledgments}
We would like to thank Miguel Escudero Abenza for valuable discussions on cosmology and neutrinos. We are grateful to Giovanni Benato, Kunio Inoue, and Andrea Pocar --- as well as to the CUORE, EXO-200, GERDA, and KamLAND-Zen Collaborations --- for providing the likelihood functions. M.A. thanks Giovanni Benato, Jason Detwiler, Javier Men\'{e}ndez and Francesco Vissani for valuable discussions.
This work has been supported by the Science and Technology Facilities Council, part of
U.K. Research and Innovation (Grant No. ST/T004169/1), the University College London (UCL) Cosmoparticle Initiative, and by the Deutsche Forschungsgemeinschaft (DFG, German Research Foundation) under the Sonderforschungsbereich (Collaborative Research Center) SFB1258 ‘Neutrinos and Dark Matter in Astro- and Particle Physics’.

\section{Appendix}
We present here the calculational procedure for the discovery probability.
As a first step, we define two hypotheses: 
\begin{itemize}
\item[$H_0$:] \nldbd is not present and all counts are background
\item[$H_1$:]  \nldbd exists and can provide event counts in addition to the background.
\end{itemize}
The probability of the hypothesis given the data is contained in the posterior probabilities $P(H_0|D)$ and $P(H_1|D)$. If the probability of $H_1$ is larger than $H_0$ then a discovery can be claimed.  For this purpose, we calculate the posterior odds
\begin{equation}
    \mathcal{O}_1 = \frac{P(H_1|D)}{P(H_0|D)}.
\end{equation}
We define a discovery by $\mathcal{O}_1 > 10$. Assuming that the two hypotheses are exhaustive we can calculate the posterior probabilities with
\begin{equation}
    P(H_i|D)= \frac{P(D|H_i)P(H_i)}{P(D|H_1)P_0(H_1)+P(D|H_0)P(H_0)},
\end{equation}
yielding
\begin{equation}
    \mathcal{O}_1 = \frac{P(D|H_1)}{P(D|H_0)} \frac{P(H_1)}{P(H_0)}.
\end{equation}
We set the prior odds $P(H_1)/P(H_0) = 1$ implying that we take the scenarios to be equally probable. 
We model the background likelihoods with Poisson distributions
\begin{equation}
    P(D|H_0) = P(\{n\}|H_0) = \prod_i e^{-\lambda_i}\frac{\lambda_i^{n_i}}{n_i!},
    \label{background}
\end{equation}
where the $\lambda_i$'s are the background expectation, which is given for each experimental setup individually, $i$ runs over the number of experiments and $\{n\}$ is the collection for the counts reported by the experiments. 
For hypothesis $H_1$, one has to add the signal expectation $\nu_i$ which are related to the experimental setups by
\begin{equation}
    \nu_i = \frac{N_A\,ln2}{m_{i}}\,\frac{\mathcal{E}_i \,\epsilon_i}{T_{1/2}},
    \label{signalexpectation}
\end{equation}
with the Avogadro number $N_A$, the molar mass of the enriched isotope $m_{i}$, the exposure $\mathcal{E}_i$ and the detection efficiency $\epsilon_i$ of the experiments, yielding 
\begin{equation}
    P(\{n\}| \theta, H_1) = P(\{n\}|\nu(\theta),H_1) = \prod_i e^{-(\lambda_i + \nu_i)}\frac{(\lambda_i + \nu_i)^{n_i}}{n_i!},
\end{equation}
where $\theta$ is the collection of parameters relevant for a \nldbd decay discussed in the text. With these definitions we can calculate $P(D|H_1)$ via
\begin{align}
 P(D|H_1) =   P(\{n\}|H_1) = \int_0^{\infty} P(\{n\}|\nu(\theta)) P(\theta|H_1) d\theta \\
 = E(P(\{n\}|\nu(\theta)))_{P(\theta)}.
 \label{signal}
\end{align}

With the quantities given in equations (\ref{background}) and (\ref{signal}) we calculate the posterior odds for a given data set. To calculate the discovery probability $P_D$, we have to create samples of possible counts the different experiments could report. We also need to sample over the possible parameter values from the analysis of available data.
The resulting mathematical expression is
\begin{equation}
    P_D = E\left[E\left[I\left(\frac{E[P(\{n\}|\theta)]_{P(\theta)}}{P(\{n\}|H_0)}\right)\right]_{P(\{n\}|\theta)}\right]_{P(\theta)}.
    \label{DiscoveryProb}
\end{equation}

Technically we use $3000$ Markov chain Monte Carlo samples from the posterior probability distribution of the analysis of available data for each investigated scenario (see \figurename~\ref{Scenarios}). Then we create $3000$ samples from the investigated experiments for each of these parameter sets. In the last step, we average over the parameter samples again while keeping the specific set of counts fixed.
By calculating the posterior odds via (\ref{BayesFactor}) for every single created event, we can decide if this specific sample we call a discovery or not and evaluate how many of the
investigated samples lead to a discovery. This procedure was followed for the single experiment case already in \cite{Caldwell:2017mqu}.

\bibliography{literatur.bib}
\end{document}